\documentclass[twocolumn,prb,showkeys,graphicx,superscriptaddress]{revtex4}
\usepackage{graphicx}
\usepackage{dcolumn}
\usepackage{bm}
\usepackage[cmex10]{amsmath}
\usepackage[utf8]{inputenc}
\usepackage{float}
\usepackage{amssymb}
\usepackage[T1]{fontenc}
\usepackage{natbib,hyperref}
\usepackage{color,soul}

\usepackage[utf8]{inputenc}
\usepackage{textcomp}
\usepackage{palatino}

\usepackage{graphicx}
\usepackage{epsfig}

\begin{document}

\preprint{APS/123-QED}

\title{Electric-field tunable spin diode FMR in patterned PMN-PT/NiFe structures
}

\author{Slawomir Zi\k{e}tek}
\email{zietek@agh.edu.pl}
\affiliation{AGH University of Science and Technology, Department of Electronics, Al. Mickiewicza 30, 30-059 Krak\'{o}w, Poland}
\author{Piotr Ogrodnik}
\email{piotrogr@if.pw.edu.pl}
\affiliation{AGH University of Science and Technology, Department of Electronics, Al. Mickiewicza 30, 30-059 Krak\'{o}w, Poland}
\affiliation{Warsaw University of Technology, Faculty of Physics, , ul. Koszykowa 75, 00-662 Warszawa, Poland}
\author{Witold Skowroński}
\affiliation{AGH University of Science and Technology, Department of Electronics, Al. Mickiewicza 30, 30-059 Krak\'{o}w, Poland}
\author{Feliks Stobiecki}
\affiliation{Institute of Molecular Physics, Polish Academy of Sciences, ul. Smoluchowskiego 17, 60-179 Poznań, Poland}
\author{Sebastiaan van Dijken}
\affiliation{NanoSpin, Department of Applied Physics, Aalto University School of Science, P.O. Box 15100, FI-00076 Aalto, Finland}
\author{J\'{o}zef Barna\'{s}}
\affiliation{Adam Mickiewicz University, Faculty of Physics, ul. Umultowska 85, 61-614 Poznań, Poland}
\affiliation{Institute of Molecular Physics, Polish Academy of Sciences, ul. Smoluchowskiego 17, 60-179 Poznań, Poland}
\author{Tomasz Stobiecki}
\affiliation{AGH University of Science and Technology, Department of Electronics, Al. Mickiewicza 30, 30-059 Krak\'{o}w, Poland}

\date{\today}

\begin {abstract}

Dynamic properties of NiFe thin films on PMN-PT piezoelectric substrate are investigated using the spin-diode method. Ferromagnetic resonance (FMR) spectra of microstrips with varying width are measured as a function of magnetic field and frequency. The FMR frequency is shown to depend on the electric field applied across the substrate, which induces strain in the NiFe layer. Electric field tunability of up to 100 MHz per 1 kV/cm is achieved. An analytical model based on  total energy minimization and the LLG equation, with magnetostriction effect taken into account, is developed to explain the measured dynamics. Based on this model, conditions for strong electric-field tunable spin diode FMR in patterned NiFe/PMN-PT structures are derived.

\end{abstract}

\maketitle

Electric field control of magnetism at room temperature can lead to the development of efficient and low-power memories,\cite{bibes2008towards,scott2007data,li2010magnetoelectric} magnetic field sensors\cite{skowronski2012magnetic}, voltage-tunable microwave filters,\cite{tatarenko2006magnetoelectric} and oscillators.\cite{useinov2015ferroelectric} Application of  multiferroic materials can additionally lead to the design of  new electronic devices in which both the electron spin and charge are affected by an external electric field. Voltage control of magnetic anisotropy (VCMA) in multiferroics can be realized by strain transfer from a ferroelectric or piezoelectric layer to a ferromagnetic film, as the deformation of the ferromagnet changes  the magnetoelastic anisotropy {\it via} inverse magnetostriction.\cite{lahtinen2011pattern,lahtinen2012electric,streubel2013strain,franke2015reversible,buzzi2013single,cherepov2014electric,nan2014quantification,    wu2011giant,liu2013voltage,zhang2014giant,yang2014non,liu2009giant,lou2009giant,liu2011electric,liu2011tunable,liu2013voltage2,li2014driving}

Recently, the influence of electric-field induced strain on magnetic anisotropy has been demonstrated in a variety of unpatterned and patterned multiferroic heterostructures including BaTiO$_3$/FM \cite{lahtinen2011pattern,lahtinen2012electric,streubel2013strain,franke2015reversible}, PMN-PT/FM \cite{liu2009giant,wu2011giant,buzzi2013single,cherepov2014electric,liu2013voltage,nan2014quantification,zhang2014giant,yang2014non} and PZN-PT/FM \cite{lou2009giant,liu2009giant,liu2011tunable,liu2011electric,liu2013voltage2,li2014driving} 
with FM = Ni \cite{streubel2013strain,wu2011giant,buzzi2013single}, NiFe \cite{liu2011tunable}, NiCo \cite{liu2011electric}, Co \cite{yang2014non}, CoFe \cite{lahtinen2011pattern, lahtinen2012electric}, CoFeB \cite{liu2013voltage, li2014driving,zhang2014giant}, Fe \cite{franke2015reversible}, FeGaB \cite{lou2009giant,liu2013voltage2}, and Fe$_3$O$_4$ \cite{liu2009giant}. Electric-field tuning of ferromagnetic resonance (FMR) has also been studied. In most reports, strong tuning of FMR in continuous ferromagnetic films on piezoelectric substrates is inferred from microwave cavity or vector network analyzer FMR measurements \cite{liu2009giant, lou2009giant, liu2011tunable,  liu2011electric, liu2013voltage2, liu2013voltage, li2014driving}. 

	In this letter, we experimentally study electric-field tuning of FMR in patterned Ni$_{80}$Fe$_{20}$ microstrips on PMN-PT substrates using a spin diode (SD) measurement technique. Patterning of the ferromagnetic film is anticipated to introduce a magnetostatic shape anisotropy, which competes with the magnetoelastic anisotropy that is induced via transfer of piezoelectric strain. To systematically study this effect, we consider NiFe microstrips of different width. An analytical model for electric-field tunable microwave signals in confined ferromagnetic geometries is also presented. Using this model, we derive a phase diagram of the FMR frequency shift as a function of microstrip width and magnetic field strength.

On a polished PMN-PT (Pb(Mg$_{1/3}$Nb$_{2/3}$)O$_3$- –PbTiO$_3$) (011)-oriented piezoelectric substrate, a 20 nm thick layer of Ni$_{80}$Fe$_{20}$  was deposited using magnetron sputtering.  The bottom side of the crystal was covered by a 5 nm Ti/50 nm Au layer, in order to apply high voltage perpendicular to the piezoelectric substrate.
 Afterwards, NiFe microstrips of 1.5,  2.6,  6.7  \textmu m width and 90 \textmu m length along the [01-1] direction of the PMN-PT crystal were fabricated using electron beam lithography and ion-beam etching. The sample was vacuum annealed at 330$^{\circ}$C in an in-plane magnetic field to increase the anisotropic magnetoresistance (AMR) ratio. The dc resistance of the respective strips was: 842 $\Omega$, 576 $\Omega$, 201.5 $\Omega$.

A radio frequency (rf) current of amplitude $I$, flowing through the NiFe strips deposited on the PMN-PT substrate, generates a time-dependent spin transfer torque (STT) and Oersted field. These effects lead to magnetization dynamics and, because of AMR, resistance oscillations. Mixing of the oscillating current and resistance generates a dc SD voltage $V_{dc}$. The amplitude of this signal is proportional to the real part $\delta R$ of the complex amplitude of the resistance change,\cite{tulapurkar2005spin,yamaguchi2007rectification,zietek2015rectification}
$ V_{dc} = \frac{1}{2} I\, \delta R$.
The AMR effect is given by: $ R = R_\perp +\Delta R \cos^2\theta$, where
$\Delta R = R_\parallel - R_\perp$ and $R_\perp$ ($R_\parallel$) denote the resistance of the strip when the magnetization is perpendicular (parallel) to the current. The change in resistance due to a small change in the angle $\theta$ between magnetic moment and current is thus given by
$\delta R = 2 \Delta R \sin\theta \cos\theta \;\delta\theta$.
From this, one finds
$V_{dc} = I \Delta R \sin\theta \cos\theta \;\delta\theta$.
We note that $\delta\theta$ represents the real part of the complex angular changes.

\begin{figure}[h!]
\begin{center}
	\includegraphics[width=8.2cm]{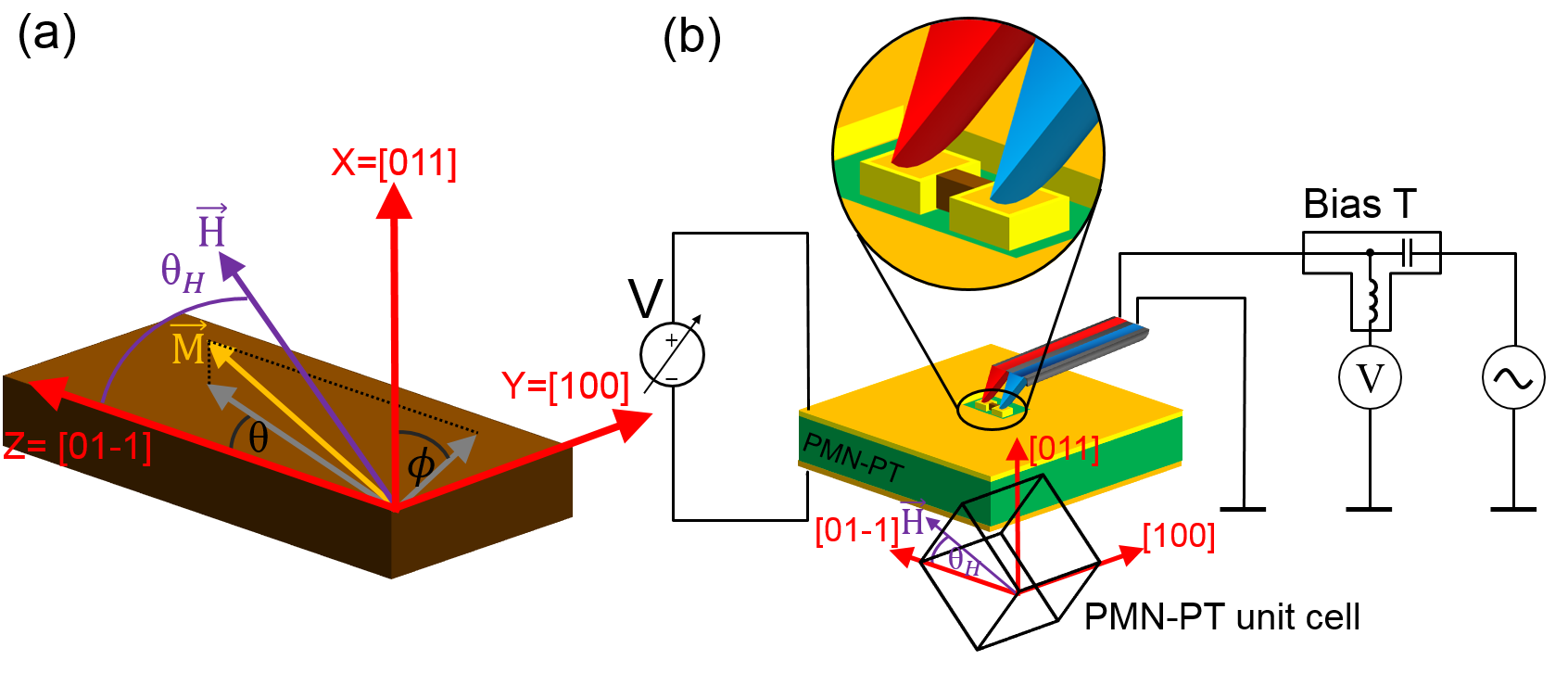}
	\caption{(a) Model geometry: the rf current is flowing along the z direction, $\theta_H$ is the angle of the external magnetic field and $\theta$ and $\phi$ denote the polar and azimuthal angles of the magnetization direction. (b) Experimental setup for SD effect measurements with external voltage applied to the PMN-PT piezoelectric crystal to electrically strain the NiFe microstrips.}
\label{fig:schemat2}
	\end{center}
\end{figure}

As mentioned above, the magnetization dynamics responsible for the
$V_{dc}$
signal is driven by the uncompensated Oersted field and by the STT effect.
Even for a single-layer ferromagnetic strip, an uncompensated Oersted field can be induced, since electron scattering processes at both interfaces are generally different. The STT, on the other hand, may appear due to some inhomogeneities of the magnetization distribution within the permalloy strip.\cite{yamaguchi2007rectification} Small changes of the angle
$\theta$
can be derived from the Landau-Lifshitz-Gilbert (LLG) equation for a unit vector along the magnetization $\vec{m} =\vec{M}/M$:
\begin{equation}
\label{eq:llg}
\frac{\partial \vec{m}(\vec{r})}{\partial t} -\alpha \vec{m}(\vec{r}) \times \frac{\partial \vec{m}(\vec{r})}{\partial t} = \vec{\Gamma}
\end{equation}
where $\alpha$ is a damping constant (of the order of $10^{-3}$), and
\begin{equation}
\label{eq:gama}
 \vec{\Gamma} = -\gamma \vec{m}(\vec{r}) \times \nabla_{\vec{M}}
 \mathcal{U} (\vec{r})
 - (\vec{u}(\vec{r})\cdot \nabla)\vec{m}(\vec{r})
 \end{equation}
is a torque acting on the magnetic moment $\vec{m}$, with $\gamma $ being the gyromagnetic ratio.
The second term in Eq. \ref{eq:gama} corresponds to the STT effect.
Its amplitude, $\vec{u}(\vec{r})$, is proportional to the
magnitude of the rf current density $\vec{j}$ and its spin polarization P, i.e. $\vec{u}(\vec{r})$ $\propto \vec{j} P$.\cite{yamaguchi2009anomalous} We have omitted here the non-adiabatic term in STT, as its amplitude is usually much smaller than that of the adiabatic term\cite{thiaville2008electrical}.
The averaged magnetic energy $U=\langle \mathcal{U} (\vec{r} )\rangle$ includes the shape anisotropy, Zeeman-like terms due to static
and dynamic (Oersted)
magnetic fields, as well as terms related to the stress due to the deformation of the PMN-PT substrate under external electric field. The energy $U$
can be written as:
\begin{eqnarray}
\label{eq:energy}
U = -  \vec{M}\cdot \vec{H}_d - \vec{M}\cdot\vec{H}-\frac{3}{2} \lambda \sigma_{[01-1]}(E) \cos^2 \theta
\nonumber \\
-[K_{[100]}- \frac{3}{2} \lambda \sigma_{[100]}(E)]\sin^2 \theta\sin^2\phi
-\vec{M}\cdot\vec{H}_{Oe[100,011]},
\end{eqnarray}
where $\vec{H_d}$ is the demagnetizing field, $\vec{H}$ is the external magnetic field, and $K_{[100]}$ is the uniaxial magnetocrystalline anisotropy that is induced during magnetic-field annealing. The coordinate system with defined angles describing orientation of the magnetization ($\theta$) and magnetic field ($\theta_H$)  with respect to the crystallographic axes of the PMN-PT crystal is shown in Fig. \ref{fig:schemat2}(a). The easy axis in our sample is transverse to the strip long axis, i.e. along the [100] direction.
In turn, $\vec{H}_{Oe[100,011]}$ in Eq. 3  denotes the time-dependent Oersted field components in the [100] or [011] directions. Because its amplitudes are constant within the sample, the Oersted field  may be written as a gradient with respect to $\vec{M}$ of the energy term $-\vec{M}\cdot\vec{H}_{Oe[100,011]} $.
Finally, $\lambda$ in Eq. 3. is the magnetostriction constant. The electric field-dependent stress $\sigma(E)$ acting on permalloy in the [100] and [01-1] directions can be calculated from the relation between the strain of the PMN-PT substrate and the applied electric field,
$\epsilon_i = d_{ji} E_j$,
where $\epsilon_i$ denotes the strain of the PMN-PT along the $i$ direction ($i = (Y,Z) \equiv [100],[01-1]$), $d_{ji}$ is a matrix of piezoelectric constants, and $E_j$ is the electric field applied in the  $j$ direction ($j = X \equiv  [011]$). Since the electric field is applied perpendicularly to the substrate (in [011] direction), the only strains that influence the permalloy microstrip are those in [100] and [01-1] directions. In particular, the relation between the strain within PMN-PT and stress transmitted to permalloy can be written as:\cite{nan2014quantification}
\begin{subequations}
\begin{align}
\label{eq:stress}
\sigma_{[01-1]}(E) = \frac{Y(d_{31}+\nu d_{32})} {(1-\nu^2)} E_{[011]}, \\
\sigma_{[100]}(E) = \frac{Y(d_{32}+\nu d_{31})} {(1-\nu^2)} E_{[011]},
\end{align}
\end{subequations}
where Y=200 GPa is the Young's modulus, and $\nu=0.3$ is Poisson's ratio of NiFe.\cite{levy2001stern}
From the above equations one can see that stresses in both in-plane directions ([100] and [01-1]) of permalloy may be different. Thus, depending on their signs, they can induce either easy or hard magnetization axes along the [100] or [01-1] directions of the NiFe microstrips. In our case,  $d_{33}$ equals $1740\pm 91\frac{pC}{N}$ (discussed in experimental part), which agrees well with data reported by M. Shanthi et al. in Ref.~\onlinecite{shanthi2008complete} for a PMN-PT crystal. Based on this, we assume the remaining piezoelectric coefficients: $d_{31}\equiv d_{[011][01-1]}=723\pm 20 \frac{pC}{N}$ and $d_{32}\equiv d_{[011][100]}=-1761\pm 13 \frac{pC}{N}$. As a consequence, the strain-induced easy magnetization axis is aligned along the [01-1] direction, which is parallel to the magnetostatic  shape anisotropy of the microstrip.

From the LLG equation one finds $\delta\theta$ as a function of the driving frequency $f$. The solution has a general resonance-curve form:\cite{zietek2015rectification} 
\begin{equation}
\delta \theta = \frac{\cos\Psi [A\sigma f^2 + B(f^2-f_0^2)] + \sin\Psi [ Af (f^2 - f_0^2) - B\sigma f ]}{(f^2 - f_0^2)^2 +\sigma^2 f^2}
\end{equation}
where A and B describe the amplitudes of the symmetric and antisymmetric contributions to the $V_{dc}$ signal, while  $f_0$ denotes the resonance frequency given by:
\begin{widetext}
\begin{equation}
f_0 \equiv  = \frac{1}{2\pi}\frac{\gamma}{(1+\alpha^2)}
\sqrt{
\begin{array}{c}
\left( \frac{\partial ^2 U}{\partial \phi^2}\frac{\partial^2 U}{\partial \theta^2} - \left[ \frac{\partial^2 U}{\partial\phi\partial\theta} \right]^2  \right)(1+\alpha^2)\csc^2\theta  + \cot\theta \left(  \frac{\partial U}{\partial \theta} \frac{\partial^2 U}{\partial \phi \theta}\alpha \csc\theta   \right. \\
\left. + \left( 2 \frac{\partial U}{\partial \phi} \frac{\partial^2 U}{\partial \theta \partial \phi}\alpha^2 + \frac{\partial U}{\partial \phi}\frac{\partial^2 U}{\partial \phi \partial \theta } - \frac{\partial U}{\partial \theta} \frac{\partial ^2 U}{\partial \phi^2}  \right)\csc^2\theta - \frac{\partial U}{\partial \theta}\frac{\partial^2U}{\partial \phi^2} \alpha \csc^3 \theta \right)
\end{array}
}
\label{eq:disp_gen}
\end{equation}
\end{widetext}
with the partial derivatives calculated at the stationary angles ($\theta,\phi$) determined by minimum magnetic energy (Eq. (\ref{eq:energy})) in the absence of rf current. Because of lack of dc current, the resonance frequency does not depend on the STT- and Oersted-field-related terms. On the other hand, the resonance frequency implicitly depends on applied electric field due to the presence of static electric-field related energy terms.

\begin{figure}[h!]
\begin{center}
	\includegraphics[width=7.8cm]{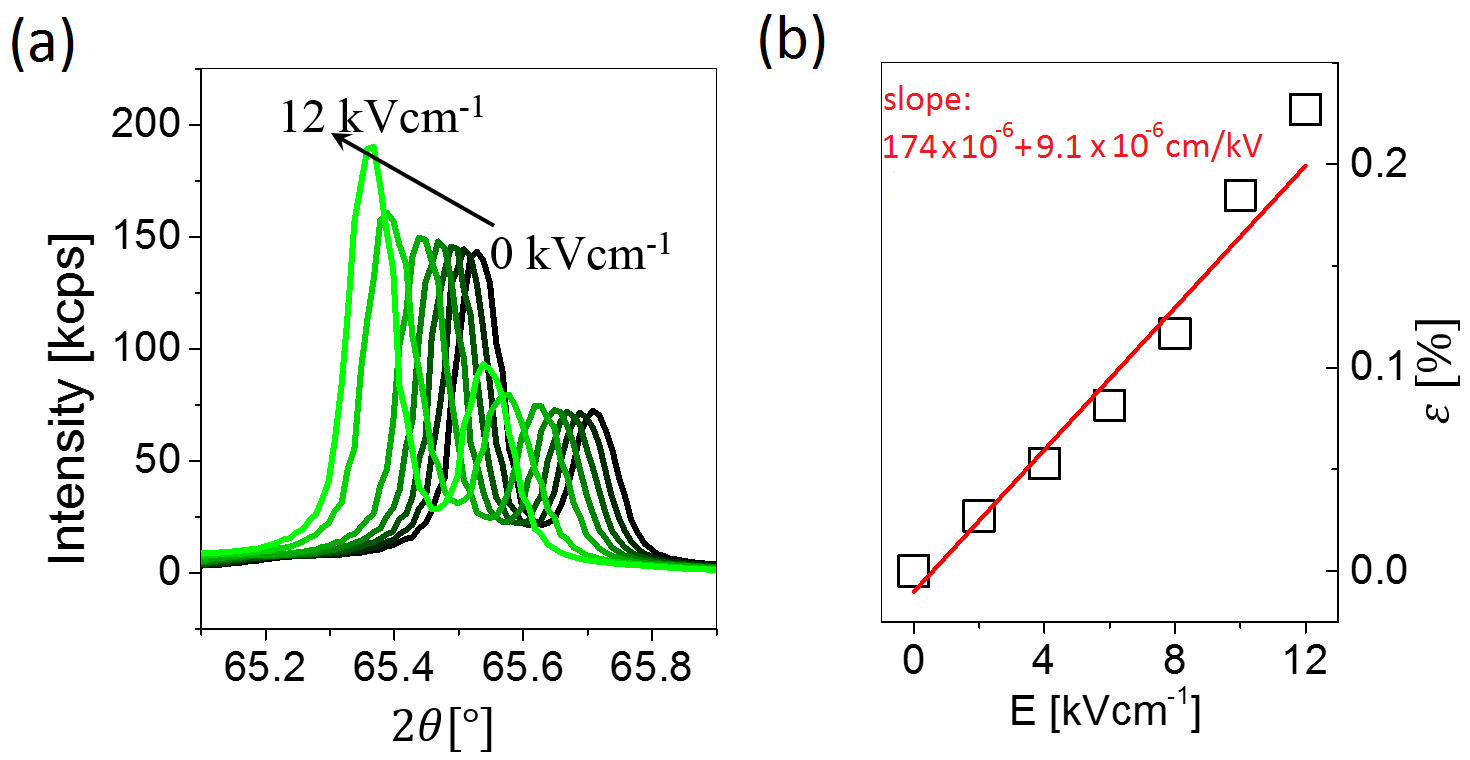}
	\caption{(a) XRD $\theta$-2$\theta$ scans of the (022) reflection of the PMN-PT substrate for electric fields  ranging from 0 to 12 kV/cm. (b) Electric field induced relative change of the out-of-plane (011) lattice parameters of the PMN-PT substrate. The red line represents a linear fit. 
}
\label{fig:XRD}
	\end{center}
\end{figure}

A schematic of the experimental setup for SD measurements of the FMR effect is presented in Fig. \ref{fig:schemat2}(b). A microwave signal of 10 dBm was applied to the NiFe microstrip using a rf probe and generator, and the dc voltage  produced by mixing of the rf current with resistance oscillations was detected by a dc voltmeter. 
Experiments with unpatterned PMN-PT/NiFe sample were also conducted as reference. X-ray diffraction (XRD) scans of the (022) PMN-PT reflection were measured in an electric field  ranging form 0 to 12 kV/cm (Fig. \ref{fig:XRD}(a)). The electric-field induced piezoelectric strain $\epsilon$ that is derived from these measurements is summarized in Fig. \ref{fig:XRD}(b). The slope of the curve, which corresponds to the $d_{33}$ piezoelectric constant, equals $174\times10^{-6}\pm 9.1$ cm/kV.

\begin{figure}[h!]
\begin{center}
	\includegraphics[width=8.4cm]{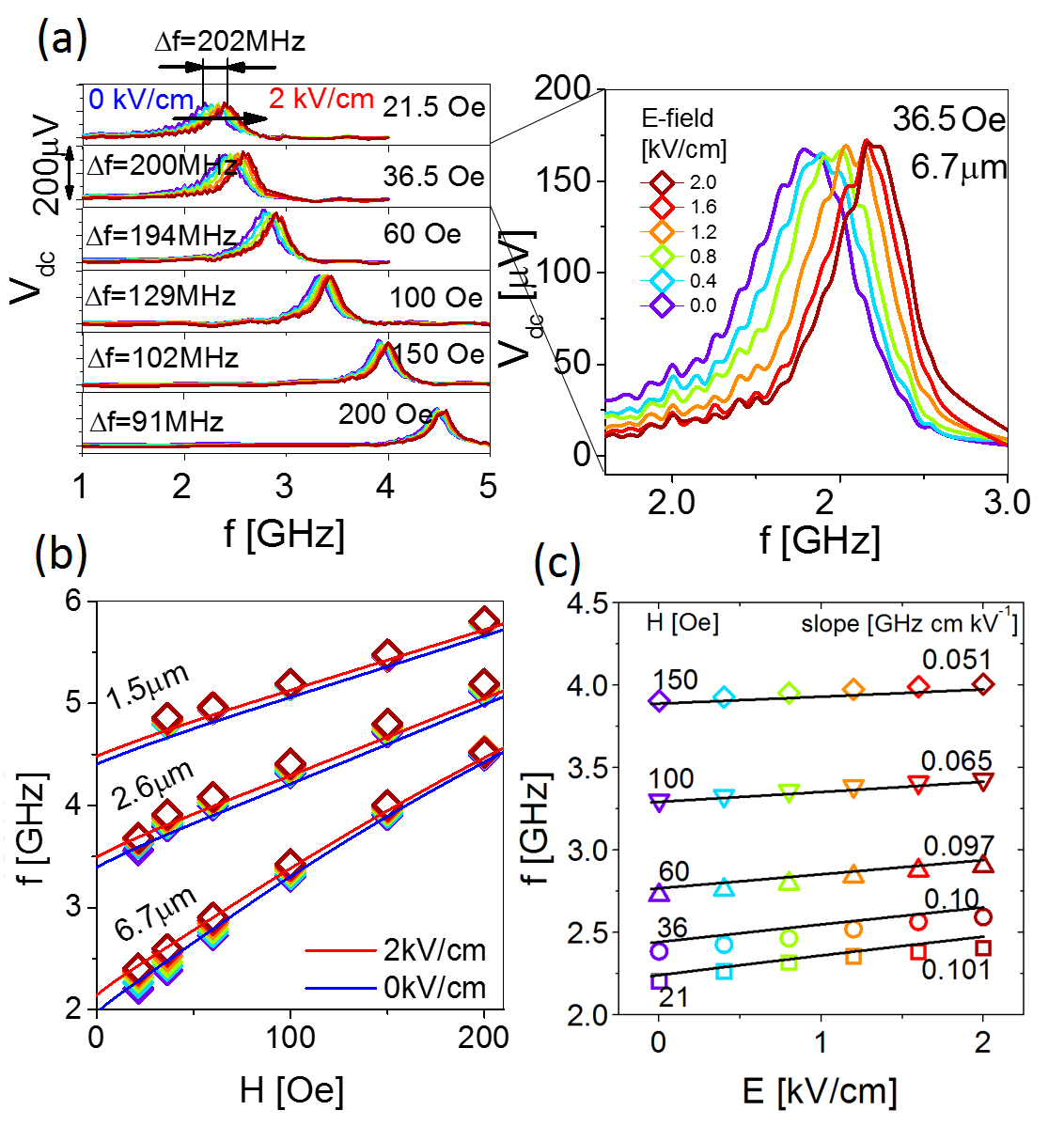}
	\caption{(a) Electric-field controlled FMR spectra at constant magnetic fields (in the range of 20-200 Oe) as a function of frequency for NiFe microstrips with a width of 6.7 \textmu m. The resonance curves for 36.5 Oe are magnified in the right panel.  (b) Dispersion relations for three microstrips with different width measured at 0 and 2 kV/cm applied electric field. Solid lines represent theoretical predictions calculated using Eq. \ref{eq:disp_gen} with $K_{[100]}=785\frac{J}{m^3}$, $\alpha=0.007$,  $M_s=0.97T$ (determined form VSM measurements), and $\lambda_{NiFe}=2.5\times10^{-6}$. (c) FMR frequency as a function of applied electric and magnetic field for NiFe microstrips with a width of 6.7 \textmu m.}
\label{fig:dysp}
	\end{center}
\end{figure}

 We used the SD effect to characterize voltage-tunable FMR in our patterned PMN-PT/NiFe structures for dc voltages in the range of 0–100 V (0-2 kV/cm). Examples of electric-field tunable FMR spectra are shown in Fig. \ref{fig:dysp}(a). The spectra were measured in a constant magnetic field (20-500 Oe) applied at $\theta_H=40 ^{\circ}$, which corresponds to the angle for which maximum SD voltages have been observed\cite{yamaguchi2007rectification}. The evolution of FMR with applied magnetic field strength for three different NiFe strips in an electric field up to 2 kV/cm are shown in Fig. \ref{fig:dysp}(b). The largest voltage-induced shift in FMR frequency (202 MHz) is observed for the widest strip at low magnetic field (21.5 Oe). For more narrow strips and larger magnetic fields, the voltage-induced frequency shifts are smaller. Fig. \ref{fig:dysp}(c) shows the FMR frequency as a function of electric and magnetic fields obtained for the widest strip, compared with theoretical model calculations (solid lines).

\begin{figure}[h!]
\begin{center}
	\includegraphics[width=9cm]{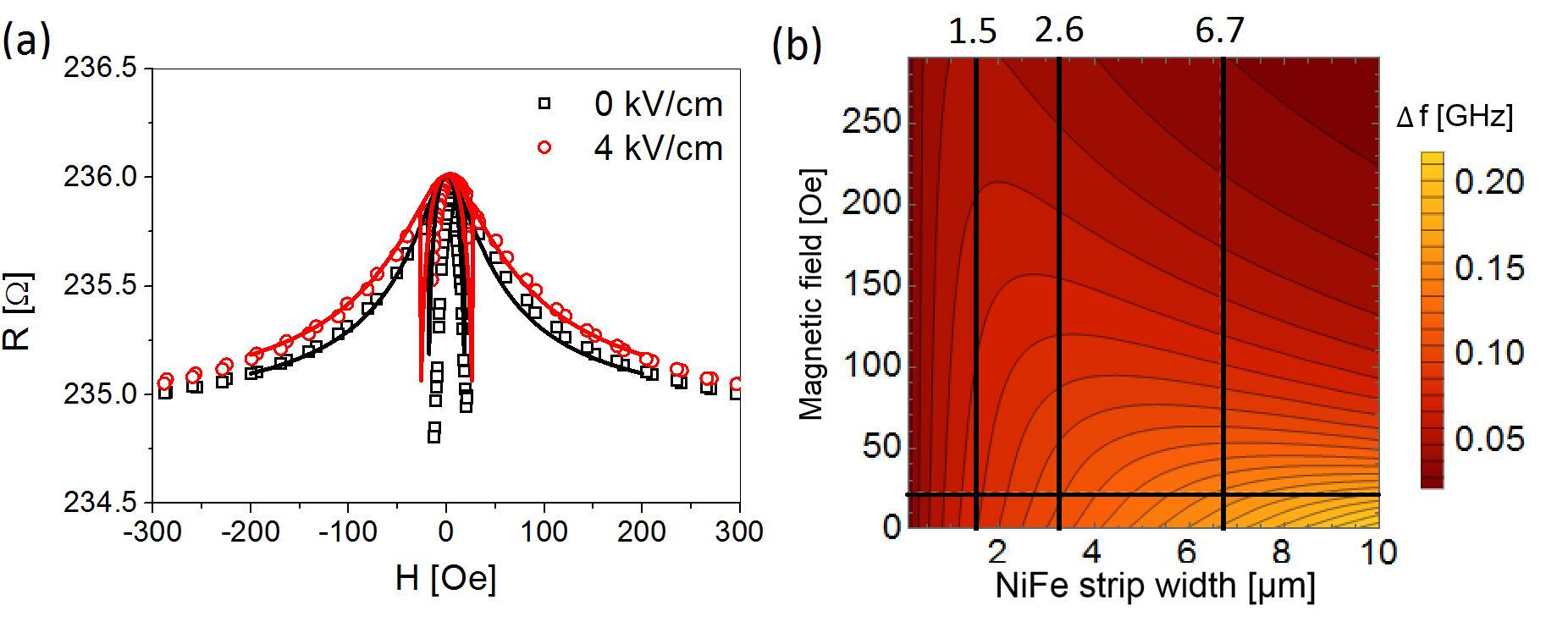}
	\caption{(a) The electric field effect on the $R(H)$ loop measured at  $\theta_H=40^{\circ}$. The solid lines represent the macrospin calculations. (b) Calculated shift of the resonance frequency, $\Delta f$, as a function of the external magnetic field and strip's width.}
\label{fig:RH}
	\end{center}
\end{figure}

To additionally proof the consistency of the derived model, AMR loops on 6.7 \textmu m wide strips were measured (Fig. \ref{fig:RH}(a)). The theoretical curves, indicated by solid lines, were calculated for 0 kV/cm and 4 kV/cm using the same parameters as for the calculated FMR shifts. 
The small discrepancy in switching fields between the theoretical and experimental results may be caused by magnetic domain formation or thermally activated switching, which are not taken into account in the macrospin model. Apart from the hysteretic region, however, the macrospin model predictions fit the experimental data well. 

This allows us to model the frequency shift in an electric field of 2 kV/cm as a function of NiFe strip width and external magnetic field using system parameters that are derived from dynamic and static measurements. The result is shown in Fig. \ref{fig:RH}(b).
The calculated phase diagram can be used to identify the parameter space for which strong electric-field tuning of FMR can be attained in NiFe microstrips on piezoelectric PMN-PT. The results indicate that the largest effects are obtained in wide microstrips at modest external magnetic field. The reduction of strain sensitivity in more narrow strips can be attributed to stronger magnetostatic shape anisotropy, which reduces the effect of the piezostrain induced magnetoelestic anisotropy.

To summarize, we have explored electric-field tuning of FMR in NiFe microstrips on the piezoelectric PMN-PT substrate. Our results indicate that electrical control of FMR spectra depends sensitively on the shape of the NiFe microstructures and applied magnetic field. A newly developed analytical model accounts for these effects and can be used to calculate FMR phase diagrams based on well-known material parameters.

\section*{Acknowledgement}

This work is supported by the Polish National Science Center grant Harmonia-DEC-2012/04/M/ST7/00799. S.v.D. acknowledges financial support from the European Research Council (ERC-2012-StG 307502-E-CONTROL). S.Z acknowledges the Dean's grant 15.11.230.198.




\end{document}